%% file: SCF420.tex
\documentclass[conference]{IEEEtran}
\IEEEoverridecommandlockouts

\usepackage[nospace]{cite}
\usepackage{amsmath,amssymb,amsfonts}
\usepackage[linesnumbered,ruled,vlined]{algorithm2e}
\usepackage{algorithmic}
\usepackage{graphicx}
\usepackage{textcomp}
\usepackage{xcolor}
\usepackage{tikz}
\usepackage{multirow}
\usepackage[caption=false]{subfig}
\usepackage{pgfplots}
\pgfplotsset{compat=1.17}
\usepackage{hyperref}
\usepackage{standalone}

\usetikzlibrary{arrows}
\usetikzlibrary{patterns}
\def\BibTeX{{\rm B\kern-.05em{\sc i\kern-.025em b}\kern-.08em
    T\kern-.1667em\lower.7ex\hbox{E}\kern-.125emX}}

\SetArgSty{textup}
\SetAlFnt{\scriptsize}

\SetCommentSty{mycommfont}

\begin{document}
\bstctlcite{IEEEexample:BSTcontrol}
\title{Lossless 4:2:0 Screen Content Coding Using Luma-Guided Soft Context Formation}
\author{\IEEEauthorblockN{Hannah Och and Andr\'{e} Kaup}
	\IEEEauthorblockA{
		Friedrich-Alexander-Universität Erlangen-Nürnberg\\
		Multimedia Communications and Signal Processing \\
		Cauerstraße 7, 91058 Erlangen, Germany}
}
\maketitle

\begin{abstract}
The soft context formation coder is a pixel-wise state-of-the-art lossless screen content coder using pattern matching and color palette coding in combination with arithmetic coding. It achieves excellent compression performance on screen content images in RGB 4:4:4 format with few distinct colors. In contrast to many other lossless compression methods, it codes entire color pixels  at once, i.e., all color components of one pixel are coded together. Consequently, it does not natively support image formats with downsampled chroma, such as YCbCr 4:2:0, which is an often used chroma format in video compression. In this paper, we extend the soft context formation coding capabilities to 4:2:0 image compression, by successively coding Y and CbCr planes based on an analysis of normalized mutual information between image planes. Additionally, we propose an enhancement to the chroma prediction based on the luminance plane. Furthermore, we propose to transmit side-information about occurring luma-chroma combinations to improve chroma probability distribution modelling. Averaged over a large screen content image dataset, our proposed method outperforms HEVC-SCC, with HEVC-SCC needing 5.66\% more bitrate compared to our method.
\end{abstract}

\begin{IEEEkeywords}
lossless coding, screen content, soft context formation, 4:2:0 coding.
\end{IEEEkeywords}

\section{Introduction}
In contrast to sensor-generated content, screen content (SC) refers to images and videos directly generated by the computer, such as web pages, presentations or graphic user interfaces of applications. In times of increasing home office usage, online conferencing and online learning opportunities, the importance of efficient compression of such data rises. Screen content has unique characteristics which differ from those of sensor-generated data. In general, SC contains only few unique colors, with large uniform areas, sharp contrasts and repetitive image areas, such as letters in textual areas. Compression methods for SC aim to exploit these characteristics.
Recent coding standards, such as High Efficiency Video Coding (HEVC) with its SC coding extension \cite{Sullivan2012,Xu2016} or Versatile Video Coding (VVC) \cite{Bross2021}, incorporate coding tools based on these characteristics to improve compression efficiency when coding SC data, such as intra block copy (IBC) or palette mode  \cite{Xu2016,Nguyen2021}.

Since distortions in synthetic image areas are more disturbing to the human visual system than distortions in natural image regions, lossless compression of SC is of high importance \cite{Peng2016}. For SC data in RGB 4:4:4 format, there exist multiple dedicated compression techniques, such as the Free Lossless Image Format (FLIF) \cite{Sneyers2016} or JPEG-LS \cite{Weinberger2000}, which are based  on pixel-wise prediction followed by entropy coding. Both recent standards HEVC and VVC are capable of lossless coding of SC in RGB 4:4:4 as well \cite{Xu2016,Nguyen2021}. All of these methods, however, are outperformed in terms of compression efficiency by soft context formation (SCF) coding \cite{Strutz2020,Och2024}, which employs pixel-wise context-based probability distribution estimation for color triplets in combination with an arithmetic coder.

Although SC data is usually captured in a 4:4:4 format, image and especially video data is often transmitted in YCbCr 4:2:0 format instead, i.e., the chroma planes are downsampled by a factor of two horizontally and vertically. Both video coding standards, HEVC and VVC, are capable of compressing YCbCr 4:2:0 data without loss, ensuring best possible reconstruction quality given the input data. Generally, they code each component separately, but may make use information from the Y plane to better compress the chroma planes, such as chroma block vectors derived from luma block vectors in IBC \cite{Xu2016a} or chroma sample prediction based on luma samples \cite{Huo2023,Li2022}.
The pixel-wise lossless compression methods do not natively support 4:2:0 formats and instead workarounds, such as separate coding of each plane, have to be employed. However, in \cite{Och2024a}, it has been shown for RGB 4:4:4 format that partial lossless coding of intra-frames using SCF can be useful for VVC even for lossy coding. For YCbCr 4:2:0 data, however, this is currently not feasible, since SCF and similar lossless coding approaches do not support the 4:2:0 chroma format. Since SCF coding has proven superior to other methods such as FLIF or JPEG-LS for RGB 4:4:4 SC data \cite{Strutz2020, Och2024}, in this paper, we focus on extending it towards 4:2:0 format capability ensuring more versatility for SCF coding and its possible use cases. 

The SCF algorithm codes an image pixel by pixel, where each color is represented as a color triplet containing all color components, e.g., red (R), green (G) and blue (B). For each pixel, the entire color triplet is coded at once. Since 4:2:0 images do not contain all three color components for each pixel, the algorithm cannot be directly extended towards 4:2:0 data.
Consequently, this paper is organized as follows: First, we will shortly recap the main points of SCF coding necessary for the following explanations. Next, we explain our coding strategy for extending SCF coding towards 4:2:0 capability by separate luma and chroma coding. We describe our proposed luma-guided enhanced chroma prediction as well as the novel luma-based chroma range coding (CRC) for improved chroma probability modelling. Finally, we will evaluate the effectiveness of our proposed coding algorithm by comparing the results of our method against VVC and HEVC lossless SC coding.
\section{Review of the Soft Context Formation Coder}
SCF  \cite{Strutz2020} compresses an image pixel by pixel in raster scan order. During compression, it gathers data about the image to estimate probability distributions for pixel colors. Given these probability distributions, the pixel colors are coded using an arithmetic coder.
\begin{figure}[t]
	\hfil\input{Images/diagram_overviewSCF}
	\vspace{-0.7em}
	\caption{\label{block_diagram}Overview of the SCF algorithm for one pixel.}
\end{figure}
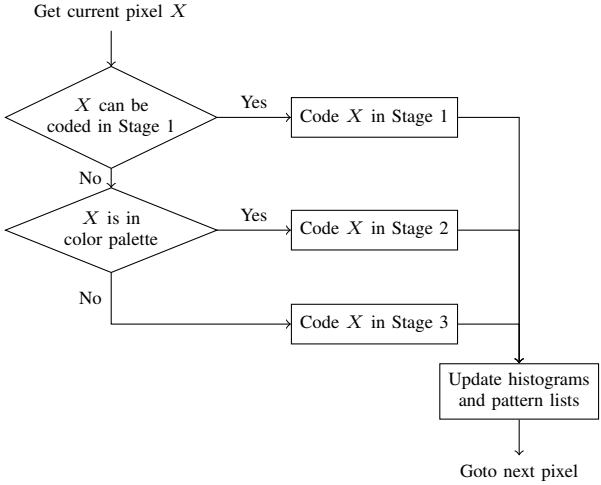
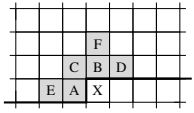
\begin{figure}[t]
	\centering
	\resizebox{0.35\columnwidth}{!}{
		\input{Images/scf_pattern.tex}
	}
	\vspace{-0.3em}
	\caption{SCF context $P=\{A,B,C,D,E,F\}$ and the corresponding current pixel at position $X$.}
	\label{fig:pattern}
	\vspace{-1em}
\end{figure}
SCF utilizes three different methods to code pixel colors: A context-based coding in Stage 1, a color-palette-based coding in Stage 2 and residual coding in Stage 3, as depicted in Figure \ref{block_diagram}. For context-based coding, the SCF algorithm saves all previously seen patterns $P = \{A,B,C,D,E,F\}$ visualized in Figure \ref{fig:pattern} in a list, and keeps an up-to-date histogram of associated colors at pixel position $X = (m,n)$. Additionally, each pattern histogram contains a count of an escape symbol, representing the number of times the histogram does not contain the color of the current pixel $X$. To code the current pixel $X$ in Stage 1, all histograms of similar patterns  to the current one are merged. The probability for a color $\mathbf{c} = [c_\mathrm{R},c_\mathrm{G},c_\mathrm{B}]^\mathrm{T}$ is then estimated as 
\begin{equation}
	p(\mathbf{c}) \approx \frac{N(\mathbf{c})}{N_P},
\end{equation}
where $N(\mathbf{c})$ is the count of the color $\mathbf{c}$ in the merged histogram and $N_P$ is the total count of all entries in the histogram. Given an arithmetic coder, we can transmit the color, needing approximately
\begin{equation}
	I(\mathbf{c}) \approx -\log_2(p(\mathbf{c}))\,\mathrm{[bit]}.
\end{equation} 
If the current color is not contained in the merged histogram, use of palette-based coding is indicated by transmitting the escape symbol instead.

For Stage 2 coding, the occurrences of each color not coded in Stage 1 are accumulated in a color palette. First, a flag indicates whether the current color is part of the palette. If so, the color of the current pixel is predicted using a modified median adaptive predictor (MAP) \cite{Strutz2020}. As described in  \cite{Strutz2020}, the color palette is split in two sub-palettes. One of them contains all colors similar to the predicted value, the other the rest. A flag indicates which sub-palette is used.  The color is then transmitted based on the distribution described by the counts of the chosen sub-palette. 

In case the current color is unknown, i.e., not yet included in the color palette, the color prediction error between actual color and MAP predicted color is transmitted per component using a histogram of previously seen prediction errors. Enhancements to this algorithm as described in \cite{Och2021, Och2024} are also applied, in addition to an image adaptive rotation \cite{Moller2019} and color transform  \cite{Strutz2020}, as well as histogram compaction.
\section{Extension of Soft Context Formation Coding For 4:2:0 Compression}
Since SCF codes entire color triplets at once, it does not natively support SC data in 4:2:0 format. Accordingly, in the following section, we will explain how to effectively extend 4:4:4 based SCF coding towards non-4:4:4 formats. 

\subsection{Separate Luma and Chroma Coding}
YCbCr is generally assumed to reduce cross-channel dependencies for natural images. In Table \ref{tab:mutual_info}, we evaluate the cross-channel dependencies for natural and SC images for both RGB and YCbCr 4:2:0. We calculate the normalized mutual information (NMI) between R-G, B-G and R-B, as well as Y-Cb, Y-Cr and Cb-Cr channels where NMI is defined as 
\begin{equation}
	\mathrm{NMI}(X_1;X_2) = \dfrac{2(H(X_1)-H(X_1|X_2))}{H(X_1)+H(X_2)},
\end{equation}
with random variables $X_1$,$X_2$, the entropy $H(\cdot)$ and the conditional entropy $H(\cdot|\cdot)$.
\begin{table}[]
	\centering
	\caption{Normalized mutual information between image channels averaged over datasets.}
	\vspace{-0.5em}
	\label{tab:mutual_info}
	\begin{tabular}{|l|ccc|ccc|}
	\hline
		& \multicolumn{3}{|c|}{RGB 4:4:4} & \multicolumn{3}{|c|}{YCbCr 4:2:0} \\
		Dataset     & R-G & R-B & G-B & Y-Cb & Y-Cr & Cb-Cr \\ \hline
		Kodak \cite{kodak} & 0.42 & 0.31 & 0.38& 0.16     & 0.14     & 0.23 \\
		SCID \cite{Ni2017}  &  0.85 & 0.73 & 0.88 & 0.32     & 0.31     & 0.50    \\ \hline
	\end{tabular}
	\vspace{-1em}
\end{table}
We average the results over the natural image dataset Kodak \cite{kodak} with 24 images and the SC image dataset SCID \cite{Ni2017} with 40 images.
To evaluate Y-Cb and Y-Cr dependencies, the chroma planes are upsampled using nearest neighbor. 

We can see that SC data shows higher mutual information between channels than the natural images from Kodak in general. The YCbCr color transform removes a lot of cross-channel dependency for both natural and SC data. However, the mutual information between Cb and Cr is still relatively high for SC images. Consequently, we adapt SCF coding for YCbCr 4:2:0 as follows: First, the adaptive color transform is no longer used since the images are already in the YCbCr color space. Second, we first code the Y channel with the SCF algorithm, i.e., each color contains only the luminance with $\mathbf{c}^\mathrm{Y}= [c_\mathrm{Y}]$. Afterwards, the combined Cb-Cr planes are coded, i.e., $\mathbf{c}^\mathrm{CbCr}= [c_\mathrm{Cb},c_\mathrm{Cr}]^\mathrm{T}$, since Cb and Cr still have a lot of interdependencies. Thus, for the Y channel, each pixel `color' contains only one component, for chroma two. Since in 4:2:0 format the chroma components are subsampled by a factor of two both horizontally and vertically, we code only one fourth the number of pixels in the chroma plane. The coder for separate Y and CbCr coding will be called `SCF 420' in the following.

\subsection{Luma-guided Prediction} 
Chroma prediction based on the luma plane has already been shown to have a positive effect \cite{Huo2023,Li2022}. However, for SC in particular, the mutual information between luma and chroma is still relatively high, see Table \ref{tab:mutual_info}. SC has often very clear structures, e.g., horizontal or vertical structures with constant colors, such as single colored text on white background. We assume that the chroma planes are similarly structured as the downsampled luma channel. Accordingly, if a pixel in the luma channel can be perfectly predicted by its top or left neighbor, the same is likely true for the chroma channel.
To find correlations between luma and chroma, we first subsample the luma plane to the same resolution as chroma, which is also done in other approaches where luma is used for chroma prediction in 4:2:0 format, e.g., in the cross-component linear model (CCLM) for chroma prediction \cite{Li2022}. 
Let $Y[m,n]$ of size $M\times N$ with spatial indices $(m,n)$ be the luminance plane of the image in YCbCr 4:2:0 format and $\tilde{Y}[m,n]$ the downsampled version of size $M/2\times N/2$.
We calculate $\tilde{Y}_s[m',n']$ at position $(m',n')$ with a precision scaling parameter $s$ as 
\begin{equation}
	\tilde{Y}_s[m',n'] = \biggl\lfloor\frac{1}{s}\left(\sum_{i=0}^{1}\sum_{j =0}^{1}Y[2m'+i,2n'+j] +\frac{s}{2}\right)\biggr\rfloor.
\end{equation}
Let $C[m',n',c]$ of size $M/2\times N/2 \times 2$ with spatial indices $(m',n')$ and color channel index  $c \in \{\mathrm{Cb},\mathrm{Cr}\}$  be the chroma planes.
For subsampling the luma plane, we use the precision scaling factor $s=2$ and enhance the MAP prediction with luma guidance according to
\begin{equation}
	\begin{aligned}
		&\mathrm{LMAP}(C[m',n',c]) = \\
		& \quad\;\;\;\begin{cases}
			C[m'-1,n',c], &  \text{if } \tilde{Y}_2[m',n'] = \tilde{Y}_2[m'-1,n'], \\
			C[m',n'-1,c], &  \text{if } \tilde{Y}_2[m',n'] = \tilde{Y}_2[m',n'-1], \\
			\mathrm{MAP}(C[m',n',c]), &  \text{otherwise},
		\end{cases}
	\end{aligned}
	\label{eq:LMAP}
\end{equation}
where  $\mathrm{MAP}(C[m',n',c])$  is the modified median adaptive prediction as described in \cite{Strutz2020}.

In Table \ref{tab:LMAP}, we show the effect of LMAP by calculating the mean absolute error (MAE) between prediction, i.e., LMAP or MAP, and true chroma planes. Furthermore, we count how often the prediction matches the actual chroma values for both Cb and Cr, i.e.,
\begin{equation}
	\mathrm{LMAP}(C[m',n',c]) = C[m',n',c]\;\;\forall c \in \{\mathrm{Cb},\mathrm{Cr}\}.
\end{equation}
We calculate the ratio of perfect predictions for all positions $(m',n')$, where one of the first two conditions of (\ref{eq:LMAP}) is met. We average the results over datasets Kodak and SCID. We can see that the prediction for SC has higher MAE values with both prediction methods than the Kodak natural image dataset, i.e., the prediction is generally worse for SC images. However, LMAP manages to improve the MAE for SC images, while the natural images remain mostly unaffected. Furthermore, with LMAP 82.7\% of Cb-Cr color doubles are correctly predicted if the luma pixel could be perfectly predicted by its top or left neighbor. 
\begin{table}[]
	\centering
	\caption{Effect of pixel-wise prediction methods MAP and LMAP on chroma planes.}
	\vspace{-1em}
	\label{tab:LMAP}
	\begin{tabular}{|l|cc|cc|}
		\hline
		& \multicolumn{2}{c|}{MAP} & \multicolumn{2}{c|}{LMAP} \\
		Dataset          & MAE     & Match in \%    & MAE     & Match in \%     \\ \hline
		Kodak \cite{kodak}& 0.708       & 42.4\%           & 0.711        & 44.3\%            \\
		SCID \cite{Ni2017}& 1.317       & 81.3\%       &  1.175      &  82.7\%               \\ \hline
	\end{tabular}
	\vspace{-1em}
\end{table}
Since we have a high percentage of perfect matches at pixels where LMAP was used, we additionally double counts of chroma values corresponding to the predicted chroma color $\mathbf{c}^\mathrm{CbCr}$ in the merged histogram of Stage 1 and the color palette in Stage 2, whereby increasing their estimated probabilities in the probability distributions.

\subsection{Luma-dependant Chroma Ranges}
The number of unique colors in SC images is often relatively small. Consequently, only few out of all possible YCbCr triplets actually occur in the image. Specifically, for each luma value in the subsampled luma image, many Y-Cb or Y-Cr combinations never happen in a specific image. At each position $(m',n')$, we evaluate the corresponding luminance value. With knowledge of which Y-Cb or Y-Cr combinations does and does not occur in the image, we can exclude chroma color doubles which cannot happen in conjunction with this luminance value from the merged histograms, the color palette or the residual coding histograms. This leads to improved compression efficiency as shown in \cite{Och2024}: Let $N_\mathrm{tot}$ be the summed counts of all entries in a histogram, e.g., the color palette, the merged histogram or residual histogram. Then, the probability of the $i$-th entry $x_i$ with count $N_i$ in the histogram is estimated as 
\begin{equation}
	p(x_i) \approx \frac{N_i}{N_\mathrm{tot}}.
\end{equation} 
If impossible entries are removed from the histogram, the new total counts of entries $N_\mathrm{tot,new}$ will be the same or smaller than before. As such, for the new estimated probability of the $i$-th entry, we have 
\begin{equation}
	p(x_i)\approx \frac{N_i}{N_\mathrm{tot,new}} \geq \frac{N_i}{N_\mathrm{tot}}.
\end{equation}
With the higher probability for the $i$-th entry, less bits are needed to transmit the corresponding symbol using an arithmetic coder.

To enable this feature, however, information about occurring Y-Cb and Y-Cr combinations has to be transmitted as side information, which can be very expensive. To reduce the number of bits necessary for this information, we use a downsampled luma image $\tilde{Y}_{s_\mathrm{CRC}}[m',n']$ with the precision scaling parameter $s_\mathrm{CRC}$, i.e., the higher the downsampling factor, the less the maximum luma value $\tilde{y}_\mathrm{max}$. Furthermore, we split the possible Cb and Cr ranges into $p$ partitions, where each partition covers a part of the entire Cb or Cr range. The $i$-th partition for Cb, for example, covers the range $\{r_\mathrm{min}(i),\dots,r_\mathrm{max}(i)\}$, where 
\begin{equation}
    \begin{aligned}
		r_\mathrm{min}(i) &= \left\lfloor \frac{i\cdot (c_\mathrm{Cb,max}+1)}{p}\right\rfloor, \\
		r_\mathrm{max}(i)& = \left\lfloor\frac{(i+1)\cdot (c_\mathrm{Cb,max}+1)}{p}\right\rfloor-1.
	\end{aligned}
\end{equation}

\begin{figure}
\centering
	\vspace{-0.4em}
	\captionsetup[subfloat]{width=0.45\columnwidth,justification=centering, singlelinecheck=false}
	\subfloat[Example image \textit{SCI12} from SCID separated into $2 \times 2$ blocks.]{
			\includegraphics[width=.44\linewidth]{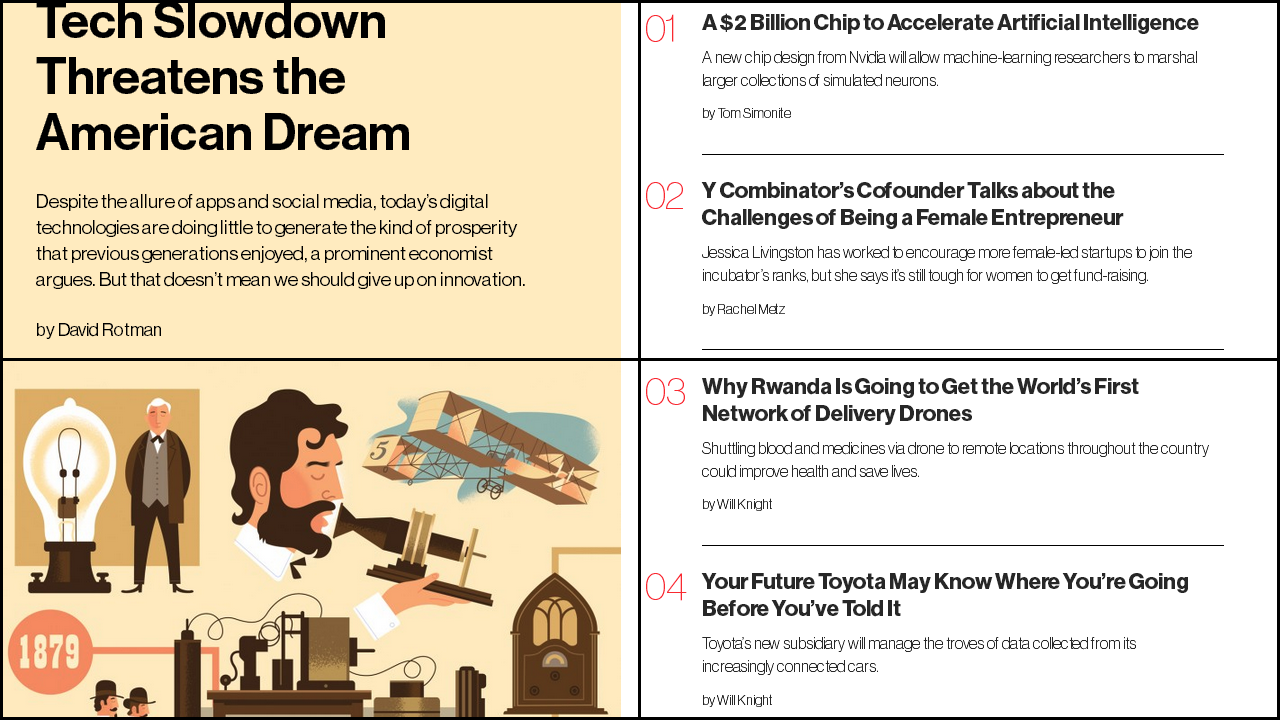}
		}
	\hspace{0.02\linewidth}
	\subfloat[ Cb range visualization\label{fig:chRange} for \textit{SCI12}.]{
			\input{Images/chRangeExample.tex}
		}
\label{fig:chRangeExmaple}
	\vspace{-0.1em}
\caption{Example of Cb range visualization for $2 \times 2$ blocks, $s = 4$ and $p=32$ for the corresponding blocks of image \textit{SCI12}.}
\end{figure}

\begin{figure}[t]
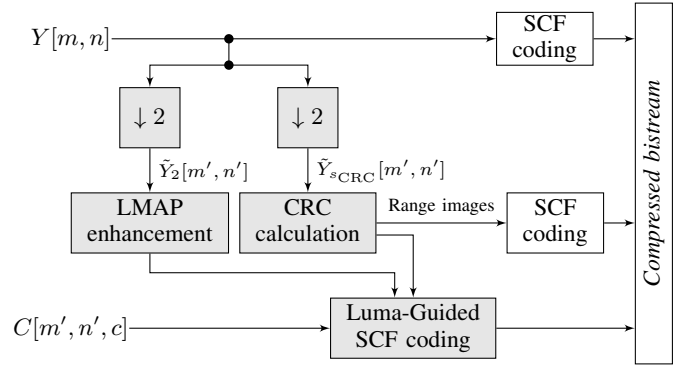

	\centering
		\vspace{-1em}
	\includestandalone{Images/diagram_420}
	\vspace{-1em}
	\caption{Overview of the proposed SCF 420 encoding pipeline. Light gray blocks indicate new blocks or modifications.}
	\label{fig:overview}
\end{figure}

Since the possible Y-Cb or Y-Cr combinations may vary strongly spatially, we also introduce a parameter $b$, which allows us to split the chroma planes into $b\times b$ blocks of equal size and to calculate the possible Y-Cb and Y-Cr combinations for each block separately. Then, for each block, we check which combinations of quantized luma sample and Cb or Cr, respectively, exist. Figure \ref{fig:chRange} shows the visualization of the Y-Cb combinations for an example image with $2\times 2$ blocks. For each block, each row in the visualization is a different $\tilde{y}$ value, with $\tilde{y} \in \left[0,\tilde{y}_\mathrm{max}\right]$. Each column depicts a partition of the Cb range. White points indicate that this combination occurs in the block, black points, that they do not occur. In the example, it is visible, that the Y-Cb combinations vary strongly depending on the block, and for many luminance values, only a small part of the possible chroma range does actually occur.
To transmit the chroma ranges, we use the original SCF algorithm and transmit both Y-Cb and Y-Cr ranges as small binary images as exemplarily shown in Figure \ref{fig:chRange}. This method will be called Chroma Range Coding `CRC' in the following.
An overview of the entire 4:2:0 coding method is visualized in Figure \ref{fig:overview}.

\section{Evaluation}
To evaluate SCF coding on 4:2:0 data, we regard five different image datasets with a total of 173 images, namely the datasets SCID \cite{Ni2017}, SIQAD \cite{Yang2015}, the `mixed' and `textual' stimuli from \cite{Shen2014}, which depict webpages, as well as the first frames from the HEVC common test conditions (CTC) dataset for SC \cite{JCTVC-U1015}. The datasets are available in RGB 4:4:4 format.  We first transform the images to YCbCr according to ITU-R BT.709 \cite{Bt.709}, and downsample the images from 4:4:4 to 4:2:0 using ffmpeg \cite{ffmpeg}. 

First, we chose optimal parameters $b$, $p$ and $s_\mathrm{CRC}$ for CRC. To this end, we evaluate the dataset SIQAD with multiple parameter combinations as visualized in Table \ref{tab:param_sweep}. We chose the parameter combination with the best average bitrate as $b=4$, $p=64$ and $s_\mathrm{CRC}= 64$.
\begin{table}[t]
	\centering
	\caption{CRC parameters obtained by training on SCID dataset \cite{Ni2017}.}
	\vspace{-0.5em}
 	\tabcolsep5pt
	\begin{tabular}{|l|c||c|}
		 \hline
		\textbf{Parameter}             & \textbf{Values}      &  \textbf{Selection} \\\hline\hline
		Number of blocks $b$               & $1,2,4$             & $4$           \\
		Precision scaling parameter $s_\mathrm{CRC}$  &$2,4,8,16,32,64,128$          & $64$           \\
		Partitions $p$         & $4,8,16,32,64,128$       & $64$           \\
		\hline
	\end{tabular}
	\vspace{-0.5em}
	\label{tab:param_sweep}
\end{table}
 \begin{table}[t]
 	\centering
 	\caption{Average bitrates in bpp for different lossless coding algorithms plus percentages w.r.t. SCF 420}
	\vspace{-0.5em}
 	\label{tab:results}
 	\begin{tabular}{|l||c|c|c|}
 		\hline
 		Dataset &
 		VTM 17.2 &
 		\begin{tabular}[c]{@{}c@{}}HM-16.21\\ SCM-8.8\end{tabular} &
 		\begin{tabular}[c]{@{}c@{}}SCF 420\\ (proposed)\end{tabular} \\ \hline \hline
 		SIQAD \cite{Yang2015}       & 2.813           & 2.715           & 2.536          \\
 		\textit{Percentage}         & \textit{110.04} & \textit{107.07} & \textit{100.00} \\ \hline
 		SCID \cite{Ni2017}          & 2.267           & 2.162           & 2.077          \\
 		\textit{Percentage}         & \textit{109.13} & \textit{104.07} & \textit{100.00} \\ \hline
 		SC-Text \cite{Shen2014}     & 1.307           & 1.262           & 1.144          \\
 		\textit{Percentage}         & \textit{114.20} & \textit{110.31} & \textit{100.00}  \\ \hline
 		SC-Mixed \cite{Shen2014}    & 1.346           & 1.329           & 1.285          \\
 		\textit{Percentage}         & \textit{104.69} & \textit{103.43} & \textit{100.00} \\ \hline
 		HEVC CTC \cite{JCTVC-U1015} & 2.136           & 2.063           & 1.968          \\
 		\textit{Percentage}         & \textit{108.53} & \textit{104.81} & \textit{100.00} \\ \hline \hline
 		All                         & 1.728           & 1.672           & 1.582          \\
 		\textit{Percentage}         & \textit{109.22} & \textit{105.66} & \textit{100.00} \\ \hline
 		\end{tabular}
			\vspace{-1em}
 \end{table}

Finally, we evaluate our resulting method in Table \ref{tab:results}. We compare our method against the lossless VVC, i.e., its reference codec VTM 17.2 \cite{VTM} with its screen content configuration in lossless mode according to the CTCs \cite{JVET_Q2014}. Additionally, we compare against HM-16.21+SCM-8.8, the reference codec for HEVC with its SC coding extension \cite{JCTVC-U1015}. Our proposed method outperforms both VVC and HEVC for every dataset, with HEVC needing between 3.43\% and 10.31\% more bitrate than our proposed method. On average, HEVC needs 5.66\% more bitrate.

To evaluate the effect of each proposed modification, we carry out an ablation study, wherein we evaluate SCF 420 without CRC and SCF 420 with neither CRC nor LMAP in Table \ref{tab:ablation}.
\begin{table}[t]
	\centering
	\caption{Ablation study: Average bitrate in bpp plus percentages w.r.t. the proposed method.}
	\vspace{-0.5em}
	\label{tab:ablation}
	\begin{tabular}{|l|c|c|c|}
		\hline
		&\begin{tabular}[c]{@{}c@{}}SCF 420\\(proposed) \end{tabular} &
		\begin{tabular}[c]{@{}c@{}}SCF 420\\ w/o CRC \end{tabular} &
		 \begin{tabular}[c]{@{}c@{}}SCF 420\\ w/o CRC and LMAP\end{tabular}\\ \hline 
		 Bitrate in bpp & 1.582         & 1.591          & 1.595            \\
		\textit{Percentage}&\textit{100.00} & \textit{100.56} & \textit{100.81} \\ \hline
	\end{tabular}
	\vspace{-2em}
\end{table}
 The SCF 420 without CRC needs 0.56\% more bitrate. Without both LMAP and CRC, SCF 420 is 0.81\% worse on average compared to our proposed method.

\begin{table}[t]
	\centering
	\caption{Average coding times in seconds.}
	\vspace{-0.5em}
	\begin{tabular}{|l|c||c|}
		\hline
      & \textbf{Enc. time}      &  \textbf{Dec. time} \\\hline\hline
      VTM 17.2          & 153.0      & 0.1          \\
      HM 21.18 SCF-8.8          & 5.8     & 0.1           \\
      SCF 420 (prop.)  & 16.8          & 15.8          \\
	\hline
	\end{tabular}
	\vspace{-1.5em}
	\label{tab:Time}
\end{table}

In Table \ref{tab:Time}, we evaluate coding times for SCF 420. We run all codecs on the same device and average the encoding and decoding times over all images. It is important to note that SCF 420 is currently not optimized for efficiency or speed. While the encoding time is still slightly slower than HEVC, it is about ten times faster than the current video coding standard VVC. Since SCF coding is a symmetric codec, it needs approximately the same time for encoding and decoding and thus its decoding times are not as fast as HM and VTM. This could be alleviated in future work by implementing a dedicated binary range image coding technique with less complexity and applying more sophisticated search algorithms for color-palette coding and context-based coding in comparison to the currently implemented sequential search.

\section{Summary}
SCF is a state-of-the-art lossless SC coding technique. Until now, it could only compress 4:4:4 content natively and was thus unsuitable for use cases where 4:2:0 compression is necessary. In this paper, we have shown that it can effectively be extended towards 4:2:0 format by separately encoding Y and CbCr planes. We further improve SCF coding for YCbCr 4:2:0 data using a luma-guided enhanced prediction for chroma samples. Additionally, we propose CRC, where we transmit knowledge about occurring luma and chroma combinations as side-information. This allows us to estimate improved color probability models for compression of the chroma planes. In total, our proposed method outperforms HEVC on all evaluated SC datasets with HEVC needing  5.66\% bitrate more on average than our proposed method.

%
%
%
\bibliographystyle{IEEEtran}
\bibliography{literature_jabRef}
\end{document}

%% file: Images/diagram_overviewSCF.tex
\usetikzlibrary{shapes.geometric}
\usetikzlibrary{positioning}

\tikzstyle{block} = [draw, fill=white, rectangle, minimum height=1.5em, minimum width=5em, align=center]
\tikzstyle{decision} = [draw, fill=white,diamond, aspect=2,minimum height=2em, minimum width=8em,align=center, inner sep=0pt, outer sep=0pt]
\tikzstyle{input} = [coordinate]
\tikzstyle{output} = [coordinate]
\tikzstyle{pinstyle} = [pin edge={to-,thin,black}]
\tikzstyle{pinstyle2} = [pin edge={-to,thin,black}]

\tikzset{font=\scriptsize}
	\begin{tikzpicture}
		\noindent
		\node [decision, pin={[pinstyle]above:Get current pixel $X$}, node distance=1.5cm] (stage1decision){$X$ can be\\ coded in Stage 1};
		\node [decision,below of=stage1decision, node distance=1.5cm] (stage2decision){$X$ is in\\ color palette};
		\node [block, right of=stage1decision, node distance=3.5cm] (stage1){Code $X$ in Stage 1};
		\node [block, right of=stage2decision, node distance=3.5cm] (stage2){Code $X$ in Stage 2};
		\node [block, below of=stage2, node distance=1.25cm] (stage3){Code $X$ in Stage 3};
		\node [block, below right=0.25cm and -0.25cm of stage3,pin={[pinstyle2]below:Goto next pixel}] (update){Update histograms\\and pattern lists};
		
		\draw  [->] (stage1decision) -- node [name=Yes1, midway, above] {Yes} (stage1);
		\draw  [->] (stage2decision) -- node [name=Yes2, midway, above] {Yes} (stage2);
		\draw  [->] (stage1decision) -- node [name=No1, midway, left] {No} (stage2decision);
		\draw [->] (stage2decision) |- node [name=No2, near start, left] {No} (stage3);
		\draw [->] (stage1) -| (update);
		\draw [->] (stage2) -| (update);
		\draw [->] (stage3) -| (update);
	\end{tikzpicture}

%% file: Images/scf_pattern.tex
\usetikzlibrary{matrix,calc}

\begin{tikzpicture}
	
	\matrix (m) [matrix of nodes, nodes={draw, minimum size=0.8cm, anchor=center}, column sep=-\pgflinewidth, row sep=-\pgflinewidth] {
		\node(m-2-1){}; & \node(m-2-2){}; & \node(m-2-3){}; & \node(m-2-4){}; & \node(m-2-5){}; & \node(m-2-6){}; & \node(m-2-7){}; \\
		\node(m-3-1){}; & \node(m-3-2){}; & \node(m-3-3){}; & \node(m-3-4){}; & \node(m-3-5){}; & \node(m-3-6){}; & \node(m-3-7){}; \\
		\node(m-4-1){}; & \node(m-4-2){}; & \node(m-4-3){}; & \node(m-4-4){}; & \node(m-4-5){}; & \node(m-4-6){}; & \node(m-4-7){}; \\
		\node(m-5-1){}; & \node(m-5-2){}; & \node(m-5-3){}; & \node(m-5-4){}; & \node(m-5-5){}; & \node(m-5-6){}; & \node(m-5-7){}; \\
	};
	
	\node at (m-5-4) [draw, fill=none, minimum size=0.8cm] {\Large X};
	
	\node at (m-5-3) [draw, fill=gray!30, minimum size=0.8cm] {\Large A};
	\node at (m-4-4) [draw, fill=gray!30, minimum size=0.8cm] {\Large B};
	\node at (m-4-3) [draw, fill=gray!30, minimum size=0.8cm] {\Large C};
	\node at (m-4-5) [draw, fill=gray!30, minimum size=0.8cm] {\Large D};
	\node at (m-5-2) [draw, fill=gray!30, minimum size=0.8cm] {\Large E};
	\node at (m-3-4) [draw, fill=gray!30, minimum size=0.8cm] {\Large F};
	
	\draw[line width=0.7mm] (m-5-1.south west)+(-0.4cm,0) -- (m-5-3.south east) -- (m-5-3.north east) -- (m-5-7.north east)+(0.4cm,0);
	\draw[line width=0.7mm] (m-5-7.north east)+(0.4cm,0) -- (m-5-3.north east);

	\foreach \i in {2,3,4,5} {
		\draw (m-\i-1.west)+(0,0.4cm) -- ++(-0.2cm,0.4cm);
		\draw (m-\i-7.east)+(0,0.4cm) -- ++(0.2cm,0.4cm);
	}
	\draw (m-5-7.east)+(0.2cm,-0.4cm) -- ++(-0.2cm,-0.4cm);
	\foreach \j in {1,2,3,4,5,6,7} {
		\draw (m-2-\j.north)+(0.4cm,0) -- ++(0.4cm,0.2cm);
		\draw (m-5-\j.south)+(0.4cm,0) -- ++(0.4cm,-0.2cm);
	}
	\draw (m-2-1.north)+(-0.4cm,0.2cm) -- ++(-0.4cm,-0.4cm);
	\draw (m-5-1.south)+(-0.4cm,-0.2cm) -- ++(-0.4cm,0.4cm);

\end{tikzpicture}

%% file: Images/chRangeExample.tex
%
%
\begin{tikzpicture}

\begin{axis}[%
width=.44\columnwidth,
at={(0,0)},
scale only axis,
point meta min=0,
point meta max=85,
axis on top,
xmin=-50,
xmax=640.5,
tick align=outside,
y dir=reverse,
ymin=-50,
ymax=600.5,
axis line style={draw=none},
ticks=none
]
\addplot [forget plot] graphics [xmin=0.5, xmax=640.5, ymin=0.5, ymax=600.5] {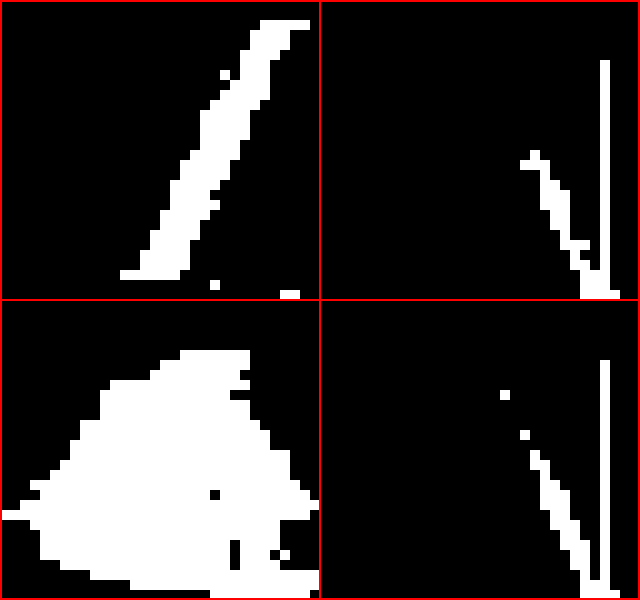};
\node[above left] at (200,20) {\tiny partitions $\rightarrow$}; 
\node[above left] at (520,20) {\tiny partitions $\rightarrow$}; 
\node[above left,rotate=90] at (15,290) { \tiny  $\leftarrow \tilde{y}$}; 
\node[above left,rotate=90] at (15,-15) { \tiny  $\leftarrow \tilde{y}$}; 
\end{axis}

\end{tikzpicture}%

%% file: Images/diagram_420.tex
	\def\radius{.7mm} 

	\begin{tikzpicture}[scale=1.4, >=latex']
		\input{Images/lib/tikzlibrarydsp}
		\small

		\tikzstyle{box} = [draw]
		\tikzset{start/.style={circle,minimum width=0.5cm,minimum height=0.3cm,draw,fill}}
		\tikzset{end/.style={draw,double=white,circle,inner sep=0pt,minimum width=0.3cm,minimum height=0.3cm,fill}}
		\node[minimum height=0.9cm,inner sep=0pt] (inY) at (0,0) {$Y[m,n]$};
		\node[minimum height=0.9cm,inner sep=0pt] (inC) at (0,-2.75) {$C[m',n',c]$};
		\node[dspnodefull] (nodey1) at (1.5,0) {};
		\node[dspnodefull] (nodey2) at (1.5,-0.25) {};
		\node[box,minimum width=0.8cm, minimum height=0.8cm,inner sep=0pt,fill=gray!20] (downsampleLMAP) at (0.75,-0.75) {$\downarrow 2$};
		\node[box,minimum width=0.8cm, minimum height=0.8cm,inner sep=0pt,fill=gray!20] (downsampleCRC) at (2.25,-0.75) {$\downarrow 2$};
		\node[box, minimum width=0.8cm, minimum height=0.8cm,inner sep=0pt,fill=gray!20] (LMAP) at (0.75,-1.75) {\begin{tabular}{c}LMAP\\enhancement\end{tabular}};
		\node[box, minimum width=0.8cm, minimum height=0.8cm,inner sep=0pt,fill=gray!20] (CRC) at (2.25,-1.75) {\begin{tabular}{c}CRC\\calculation\end{tabular}};
		\node[box, minimum width=0.8cm, minimum height=0.8,inner sep=0pt] (YSCF) at (4.5,0) {\begin{tabular}{c}SCF\\coding\end{tabular}};
		\node[box, minimum width=0.8cm, minimum height=0.8cm,inner sep=0pt,fill=gray!20] (CSCF) at (3.25,-2.75) {\begin{tabular}{c}Luma-Guided\\SCF coding\end{tabular}};
		\node[box, minimum width=0.6cm, minimum height=0.8cm,inner sep=0pt] (CRCSCF) at (4.6,-1.75) {\begin{tabular}{c}SCF\\coding\end{tabular}};
		\node[](middle) at (0,-1.375) {};
		\node [box, minimum width=4.8cm, right of=middle, rotate=90, node distance=7.75cm] (bitstream){\textit{Compressed bistream}};
		\draw[-] (inY) -- (nodey1) node [pos=0.5,above] {} node [pos=0.5,below] {};
		\draw[-] (nodey1) -- (nodey2) node [pos=0.5,above] {} node [pos=0.5,below] {};
		\draw[->] (nodey2) -| (downsampleLMAP) node [pos=0.5,above] {} node [pos=0.5,below] {};
		\draw[->] (nodey2) -| (downsampleCRC) node [pos=0.5,above] {} node [pos=0.5,below] {};
		\draw[->] (downsampleCRC) -- (CRC) node [pos=0.5,right] {\scriptsize  $\tilde{Y}_{s_\mathrm{CRC}}[m',n']$} node [pos=0.5,below] {};
		\draw[->] (downsampleLMAP) -- (LMAP) node [pos=0.5,right] {\scriptsize  $\tilde{Y}_{2}[m',n']$} node [pos=0.5,below] {};
		\draw[->] (nodey1) -- (YSCF) node [pos=0.5,above] {} node [pos=0.5,below] {};
		\draw[->] (inC) -- (CSCF) node [pos=0.5,above] {} node [pos=0.5,below] {};

		\draw[->] (LMAP) |- ([yshift=15pt] CSCF.west) -| ([xshift=-5pt] CSCF.north);
		\draw[->] ([yshift=-5pt] CRC) -| (CSCF);
		\draw[->] (CRC) -- (CRCSCF) node [pos=0.5,above] {\scriptsize  Range images};
		\draw  [->] (CSCF) -- (CSCF-|bitstream.north);
		\draw  [->] (YSCF) -- (YSCF-|bitstream.north);
		\draw  [->] (CRCSCF) -- (CRCSCF-|bitstream.north);
	\end{tikzpicture}